\documentclass{article}
\usepackage[totalwidth=480pt, totalheight=680pt]{geometry}

\usepackage{amsmath,amssymb,amsthm}
\usepackage{xspace}
\usepackage{cite}
\usepackage[scaled=.92]{helvet}
\usepackage[utf8]{inputenc}

\newtheorem{theorem}{Theorem}
\newtheorem{problem}[theorem]{Problem}
\newtheorem{definition}[theorem]{Definition}

\usepackage[usenames]{color}

\newcommand{\smartparagraph}[1]{\smallskip\noindent 
{\bf #1}\ }

\newcommand{\tool}[1]{\textsf{#1}\xspace}

\newcommand{\ifspace}[1]{}

\newcommand{\m}[1]{\mathsf{#1}}

\newcommand{\boolTrue}{\top}
\newcommand{\boolFalse}{\bot}

\newcommand{\set}[1]{\{#1\}}
\newcommand{\pset}[2]{\set{\,#1\mid#2\,}}

\def\vec#1{\mathchoice{\mbox{\boldmath$\displaystyle#1$}}
{\mbox{\boldmath$\textstyle#1$}}
{\mbox{\boldmath$\scriptstyle#1$}}
{\mbox{\boldmath$\scriptscriptstyle#1$}}}

\newcommand{\restrict}[2]{{#1}|_{#2}}

\newcommand{\abbcvc}{C}
\newcommand{\abbz}{Z3}
\newcommand{\abbnoinst}{U}
\newcommand{\abbinst}{P}
\newcommand{\abbdefault}{D}
\newcommand{\abblocal}{L}
\newcommand{\abbopt}{O}
\newcommand{\abbmbqi}{M}

\newcommand{\alg}{M}
\newcommand{\sig}{\Sigma}
\DeclareMathOperator{\sorts}{\textsf{Sorts}}
\newcommand{\funs}{\Omega}

\newcommand{\preds}{\Pi}

\newcommand{\clause}{C}

\newcommand{\support}[1]{[#1]}

\newcommand{\theory}{\mathcal{T}}

\newcommand{\weakmodels}{{\models_w}}

\newcommand{\boolsort}{\m{bool}}

\newcommand{\equal}{\approx}
\newcommand{\equalities}{E}
\newcommand{\axiom}{K}
\newcommand{\axioms}{\mathcal{K}}
\newcommand{\Theory}{\mathcal{T}}
\newcommand{\basetheory}{{\Theory_0}}
\newcommand{\exttheory}{{\Theory_1}}
\newcommand{\extaxioms}{\axioms_e}
\newcommand{\sub}{\sigma}    
\newcommand{\subs}{\mathcal{S}}   

\DeclareMathOperator*{\subterms}{\mathsf{st}}
\newcommand{\groundterms}{G}
\newcommand{\aformula}{\phi}

\newcommand{\Patterns}{P}
\newcommand{\theoryproc}{\mathfrak{D}}
\newcommand{\ematchproc}{\mathfrak{E}}

\newcommand{\ZZ}{\mathbb{Z}}

\newcommand{\varassign}{\nu}

\newcommand{\sat}{\textsf{sat}}
\newcommand{\unsat}{\textsf{unsat}}

\newcommand{\union}{\cup}

\usepackage[caption=false]{subfig}

\usepackage{eso-pic}
\usepackage{todonotes}

\newif\ifDRAFT
\newif\ifSHOWOLD
\newif\ifHIGHLIGHTNEW

\DRAFTfalse \SHOWOLDfalse \HIGHLIGHTNEWfalse

\ifSHOWOLD
\usepackage{soul}
\newcommand\kold[1]{\textcolor{gray}{\st{#1}}}
\else
\newcommand\kold[1]{}
\fi

\ifDRAFT
\newcommand\asays[1]{\todo[inline]{{\bf A says:} #1}}

\newcommand\ksays[1]{\todo[color=blue!40]{{\bf K says:} #1}}
\newcommand\ksaysi[1]{\todo[inline,color=blue!40]{{\bf K says:} #1}}

\newcommand\timsays[1]{\todo[color=red!40]{{\bf Tim says:} #1}}

\else
\newcommand\ksays[1]{}
\newcommand\ksaysi[1]{}
\fi

\ifHIGHLIGHTNEW
\newcommand\knew[1]{\textcolor{teal}{#1}}
\else
\newcommand\knew[1]{#1}
\fi

\usepackage[
  colorlinks,
  urlcolor=blue,
  citecolor=blue,
]{hyperref}

\usepackage{enumitem}

\usepackage{multirow}
\newcommand{\ColumnSpace}{@{\hskip .8ex}}
\newcommand{\ColumnBar}{\ColumnSpace|\ColumnSpace}

\usepackage{authblk}

\title{On Deciding Local Theory Extensions via E-matching}

\author[1]{Kshitij Bansal}
\author[2]{Andrew Reynolds}
\author[3]{Tim King}
\author[1]{Clark Barrett}
\author[1]{Thomas~Wies}
\affil[1]{New York University}
\affil[2]{École Polytechnique Fédérale de Lausanne}
\affil[3]{Verimag}

\date{}

\begin{document}

\maketitle
\begin{center}
\textit{To the memory of Morgan Deters}
\end{center}

\medskip
\begin{center}
  \small{
    Published in \emph{Computer Aided Verification} \\
    The final publication is available at Springer via:
    \url{http://link.springer.com/chapter/10.1007%2F978-3-319-21668-3_6}
    }
\end{center}

\begin{abstract}
  Satisfiability Modulo Theories (SMT) solvers incorporate 
  decision procedures for theories of data types that commonly occur
  in software. This makes them important tools for automating 
  verification problems.  A limitation frequently encountered is that
  verification problems are often not fully expressible in the
  theories supported natively by the solvers.
  \kold{To get around this, }Many solvers allow the specification of
  application-specific theories as quantified axioms, \kold{in their
    input. These axioms are then heuristically instantiated during the
    solver's search. Unfortunately, these heuristics are}\knew{but
    their handling is} incomplete outside of narrow special cases.
  %
  %
  %
  
  In this work, we show how SMT solvers can be used to obtain complete
  decision procedures for local theory extensions, an important class
  of theories that are decidable using finite instantiation of
  axioms. We present an algorithm that uses E-matching to generate
  instances incrementally during the search, significantly reducing
  the number of generated instances compared to eager instantiation
  strategies.  We have used two SMT solvers to implement this
  algorithm and conducted an extensive experimental evaluation on
  benchmarks derived from verification conditions for
  heap-manipulating programs. We believe that our results are of
  interest to both the users of SMT solvers as well as their
  developers.

\end{abstract}

\section{Introduction}


Satisfiability Modulo Theories (SMT) solvers are a cornerstone of
today's verification technology. Common applications of SMT include
checking verification conditions in deductive
verification~\cite{DBLP:conf/lpar/Leino10, DBLP:conf/esop/FilliatreP13},
computing program abstractions in software model
checking~\cite{DBLP:conf/fmcad/McMillan11,
  DBLP:journals/jar/BrilloutKRW11, DBLP:conf/cav/AlbarghouthiLGC12},
and synthesizing code fragments in software
synthesis~\cite{DBLP:conf/cav/BodikT12,
  DBLP:conf/popl/BeyeneCPR14}. Ultimately, all these tasks can be
reduced to satisfiability of formulas in certain first-order theories
that model the semantics of prevalent data types and software
constructs, such as integers, bitvectors, and arrays. The appeal of
SMT solvers is that they implement decision procedures for
efficiently reasoning about formulas in these theories. Thus, they can
often be used off the shelf as automated back-end solvers in
verification tools.

Some verification tasks involve reasoning about universally
quantified formulas, which goes beyond the capabilities of the
solvers' core decision procedures. Typical examples include
verification of programs with complex data structures and concurrency,
yielding formulas that quantify over unbounded sets of memory
locations or thread identifiers. From a logical perspective, these
quantified formulas can be thought of as axioms of
application-specific theories. In practice, such theories often remain
within decidable fragments of first-order
logic~\cite{DBLP:journals/jar/BrilloutKRW11,
  DBLP:conf/atva/BouajjaniDES12, DBLP:conf/tacas/AlbertiGS14,
  DBLP:conf/popl/ItzhakyBILNS14}. However, their narrow scope (which
is typically restricted to a specific program) does not justify the
implementation of a dedicated decision procedure inside the SMT
solver. Instead, many solvers allow theory axioms to be specified
directly in the input constraints. The solver then provides a
quantifier module that is designed to heuristically instantiate these
axioms.
These heuristics are in general incomplete and
the user is given little control over the instance generation.
Thus, even if there exists a finite instantiation strategy that yields a
decision procedure for a specific set of axioms,
the communication of strategies and tactics to SMT solvers is a
challenge~\cite{DBLP:conf/birthday/MouraP13}. Further,
the user cannot communicate the completeness of such a strategy.
In this situation, the user is left with two alternatives:
either she gives up on completeness, which may lead to usability
issues in the verification tool, or she implements her own
instantiation engine as a preprocessor to the SMT solver, leading to
duplication of effort and reduced solver performance.

The contributions of this paper are two-fold. First, we provide a
better understanding of how complete decision procedures for
application-specific theories can be realized with the quantifier
modules that are implemented in SMT solvers. Second, we
explore several extensions of the capabilities of these modules to
better serve the needs of verification tool developers. The focus of
our exploration is on \emph{local theory extensions}~\cite{SS05,
  IhlemannETAL08LocalReasoninginVerification}.  A theory extension
extends a given base theory with additional symbols and axioms. Local
theory extensions are a class of such extensions that can be
decided using finite quantifier instantiation of the extension
axioms. This class is attractive because it is characterized by proof
and model-theoretic properties that abstract from the intricacies of
specific quantifier instantiation techniques~\cite{G01,SS05,
  DBLP:conf/frocos/HorbachS13}. Also, many well-known theories that
are important in verification but not commonly supported by SMT
solvers are in fact local theory extensions, even if they have not
been presented as such in the literature. Examples include the array
property fragment~\cite{DBLP:conf/vmcai/BradleyMS06}, the theory of
reachability in linked lists~\cite{DBLP:conf/vmcai/RakamaricBH07,
  DBLP:conf/popl/LahiriQ08}, and the theories of finite
sets~\cite{DBLP:conf/birthday/Zarba03} and
multisets~\cite{Zarba02CombiningMultisetsIntegers}.

We present a general decision procedure for local theory extensions
that relies on E-matching, one of the core components of the
quantifier modules in SMT solvers. We have implemented our decision
procedure using the SMT solvers
\tool{CVC4}~\cite{conf/cav/BarrettCDHJKRT11} and
\tool{Z3}~\cite{MouraBjoerner08Z3} and applied it to a large set of
SMT benchmarks coming from the deductive software verification tool
\tool{GRASShopper}~\cite{DBLP:conf/cav/PiskacWZ13, grasshopper}. These
benchmarks use a hierarchical combination of local theory extensions
to encode verification conditions that express correctness properties
of programs manipulating complex heap-allocated data
structures. Guided by our experiments, we developed generic
optimizations in \tool{CVC4} that improve the performance of our
base-line decision procedure. Some of these optimizations required us
to implement extensions in the solver's quantifier module. We believe
that our results are of interest to both the users of SMT solvers as
well as their developers. For users we provide simple ways of
realizing complete decision procedures for application-specific
theories with today's SMT solvers. For developers we provide
interesting insights that can help them further improve the
completeness and performance of today's quantifier instantiation
modules.

\paragraph{Related work.}

Sofronie-Stokkermans~\cite{SS05} introduced local theory extensions
as a generalization of locality in equational
theories~\cite{DBLP:conf/kr/GivanM92, G01}. Further generalizations include Psi-local
theories~\cite{IhlemannETAL08LocalReasoninginVerification}, which can
describe arbitrary theory extensions that admit finite quantifier
instantiation. The formalization of our algorithm targets local theory
extensions, but we briefly describe how it can be generalized to
handle Psi-locality. The original decision procedure for local theory
extensions presented in~\cite{SS05}, which is implemented in
\tool{H-Pilot}~\cite{DBLP:conf/cade/IhlemannS09}, eagerly
generates all instances of extension axioms upfront, before the base
theory solver is called. As we show in our experiments, eager
instantiation is prohibitively expensive for many local theory
extensions that are of interest in verification because it results in
a high degree polynomial blowup in the problem size.

In~\cite{Jacobs09}, Swen Jacobs proposed an incremental instantiation
algorithm for local theory extensions. The algorithm is a variant of
model-based quantifier instantiation (MBQI). It uses the base theory
solver to incrementally generate partial models from which relevant
axiom instances are extracted. The algorithm was implemented as a
plug-in to \tool{Z3} and experiments showed that it helps to reduce
the overall number of axiom instances that need to be
considered. However, the benchmarks were artificially
generated. 
Jacob's algorithm is orthogonal to ours as the focus of
this paper is on how to use SMT solvers for deciding local theory
extensions without adding new substantial functionality to the solvers.
A combination with this approach is feasible as we discuss in more
detail below.

Other variants of MBQI include its use in the context of finite model
finding~\cite{ReyEtAl-CADE-13}, and the algorithm described in \cite{GM09},
which is implemented in \tool{Z3}. This algorithm is complete for the
so-called almost uninterpreted fragment of first-order logic. While
this fragment is not sufficiently expressive for the local theory
extensions that appear in our benchmarks, it includes important
fragments such as Effectively Propositional Logic (EPR). In fact, we
have also experimented with a hybrid approach that uses our
E-matching-based algorithm to reduce the benchmarks first to EPR and
then solves them with \tool{Z3}'s MBQI algorithm.

E-matching was first described in~\cite{Nelson:1980:TPV:909447},
and since has been implemented in various SMT solvers~\cite{MB07, GBT09}.
In practice, user-provided \emph{triggers} can be given as hints for
finer grained control over quantifier instantiations in these implementations.
More recent work~\cite{Dross2012} has made progress towards 
formalizing the semantics of triggers for the purposes of specifying
decision procedures for a number of theories.
A more general but incomplete technique~\cite{reynolds14quant_fmcad} 
addresses the prohibitively large number of instantiations produced by 
E-matching by prioritizing instantiations that lead to ground conflicts.

\section{Example}
\label{sec:example}

We start our discussion with a simple example that illustrates the
basic idea behind local theory extensions.
Consider the following set of ground literals
\[ G = \{ a + b = 1\text{, }f(a) + f(b) = 0 \}.
\] 
We interpret $G$ in the theory of linear integer arithmetic and a
monotonically increasing function $f:\ZZ \to \ZZ$.
One satisfying assignment for $G$ is:
\begin{equation}\label{eqn:example:model:1}
  a=0\text{, }b=1\text{, }f(x) = \{ -1\text{ if }x \leq 0, 1 \text{ if }x > 0 \}.
\end{equation}
We now explain how we can use an SMT solver to conclude that $G$ is
indeed satisfiable in the above theory.

SMT solvers commonly provide inbuilt decision procedures for common
theories such as the theory of linear integer arithmetic (\texttt{LIA})
and the theory of equality over uninterpreted functions
(\texttt{UF}). However, they do not natively support the theory of
monotone functions. The standard way to enforce $f$ to be monotonic
is to axiomatize this property,
\begin{equation}\label{eqn:example:axiom}
  \axiom = \forall x, y.\ x \leq y \implies f(x) \leq f(y), 
\end{equation}
and then let the SMT solver check if $G \cup \set{\axiom}$ is
satisfiable via a reduction to its natively supported theories. In our
example, the reduction target is the combination of \texttt{LIA}
and \texttt{UF}, which we refer to as the \emph{base theory}, 
denoted by $\basetheory$. We refer to the axiom $\axiom$ as a
\emph{theory extension} of the base theory and to the function symbol
$f$ as an \emph{extension symbol}.

Most SMT solvers divide the work of deciding ground formulas $G$ in a
base theory $\basetheory$ and axioms $\axioms$ of theory extensions 
between different modules.  A quantifier module looks for
substitutions to the variables within an axiom $\axiom$, $x$ and $y$, to some
ground terms, $t_1$ and $t_2$.  We denote such a substitution as
$\sigma = \{ x \mapsto t_1, y \mapsto t_2\}$ and the instance of
an axiom $\axiom$ with respect to this substitution as $\axiom \sigma$. The
quantifier module iteratively adds the generated ground instances
$\axiom \sigma$ as lemmas to $G$ until the base theory solver derives
a contradiction. However, if $G$ is satisfiable, as in our case, then
the quantifier module does not know when to stop generating instances
of $\axiom$, and the solver may diverge, effectively enumerating
an infinite model of $G$.

For a local theory extension, we can syntactically restrict the instances
$\axiom \sigma$ that need to be considered before concluding that $G$
is satisfiable to a finite set of candidates.  More precisely, a
theory extension is called \emph{local} if in order to decide
satisfiability of $G \cup \set{\axiom}$, it is sufficient to consider
only those instances $\axiom \sigma$ in which all ground terms already
occur in $G$ and $\axiom$. The monotonicity axiom $\axiom$ is a local
theory extension of $\basetheory$. The local instances of $\axiom$ and
$G$ are:
\begin{align*}
  \axiom \sigma_1 = a \leq b \implies f(a) \leq f(b) &
  \; \text{ where } \; \sigma_1 = \{x \mapsto a, y \mapsto b\}, \\
  \axiom \sigma_2 = b \leq a \implies f(b) \leq f(a) & 
  \; \text{ where } \; \sigma_2 = \{x \mapsto b, y \mapsto a\}, \\
  \axiom \sigma_3 = a \leq a \implies f(a) \leq f(a) & 
  \; \text{ where } \; \sigma_3 = \{x \mapsto a, y \mapsto a\}, \text{ and}\\
  \axiom \sigma_4 = b \leq b \implies f(b) \leq f(b) & 
  \; \text{ where } \; \sigma_4 = \{x \mapsto b, y \mapsto b\}.
\end{align*}
Note that we do not need to instantiate $x$ and $y$ with other ground
terms in $G$, such as $0$ and $1$. 
Adding the above instances to $G$ yields
\[ G' = G \cup \{ \axiom \sigma_1, \axiom\sigma_2, \axiom \sigma_3,
\axiom \sigma_4 \}.
\] 
which is satisfiable in the base theory. Since $\axiom$ is a local
theory extension, we can immediately conclude that $G \cup \{\axiom\}$
is also satisfiable.

\paragraph{Recognizing Local Theory Extensions.}

There are two useful characterizations of local theory extensions that
can help users of SMT solvers in designing axiomatization that are
local. The first one is model-theoretic~\cite{G01,SS05}. Consider
again the set of ground clauses $G'$. When checking satisfiability of
$G'$ in the base theory, the SMT solver may produce the following
model:
\begin{equation}\label{eqn:example:model:2}
  a=0\text{, }b=1\text{, }f(x) = \{ -1 \text{ if } x=0\text{, }1\text{ if }x=1
  \text{, -1 otherwise} \}.
\end{equation}
This is not a model of the original $G \cup \{\axiom\}$. However, if
we restrict the interpretation of the extension symbol $f$ in this
model to the ground terms in $G \cup \{\axiom\}$, we obtain the
\emph{partial model}
\begin{equation}\label{eqn:example:model:3}
  a=0\text{, }b=1\text{, }f(x) = \{ -1 \text{ if } x=0\text{, }1\text{ if }x=1
  \text{, undefined otherwise} \}.
\end{equation}
This partial model can now be embedded into the
model~(\ref{eqn:example:model:1}) of the theory extension. If such
embeddings of partial models of $G'$ to total models of $G \cup
\{\axiom\}$ always exist for all sets of ground literals $G$, then
$\axiom$ is a local theory extension of $\basetheory$. The second
characterization of local theory extensions is proof-theoretic and
states that a set of axioms is a local theory extension if it is
saturated under (ordered)
resolution~\cite{DBLP:conf/lics/BasinG96}. This characterization can
be used to automatically compute local theory extensions from
non-local ones~\cite{DBLP:conf/frocos/HorbachS13}.

Note that the locality property depends both on the base theory as
well as the specific axiomatization of the theory extension. For
example, the following axiomatization of a monotone function $f$ over
the integers, which is logically equivalent to
equation~(\ref{eqn:example:axiom}) in $\basetheory$, is not local:
\[\axiom = \forall x.\, f(x) \leq f(x + 1) \enspace. \]
Similarly, if we replace all inequalities in
equation~(\ref{eqn:example:axiom}) by strict inequalities, then the
extension is no longer local for the base theory
$\basetheory$. However, if we replace $\basetheory$ by a theory in
which $\leq$ is a dense order (such as in linear real arithmetic),
then the strict version of the monotonicity axiom is again a local
theory extension.

In the next two sections, we show how we can use the existing
technology implemented in quantifier modules of SMT solvers to decide
local theory extensions. In particular, we show how E-matching can be
used to further reduce the number of axiom instances that need to be
considered before we can conclude that a given set of ground literals
$G$ is satisfiable.

\section{Preliminaries}
\label{sec:prelim}

\kold{In the following, we define the syntax and semantics of formulas.}

\paragraph{Sorted first-order logic.}  We present our problem in
sorted first-order logic with equality. A \emph{signature} $\sig$ is a
tuple $(\sorts, \funs, \preds)$, where $\sorts$ is a countable set of
sorts and $\funs$ and $\preds$ are countable sets of function and
predicate symbols, respectively. Each function symbol $f \in \funs$
has an associated arity $n \geq 0$ and associated sort $s_1 \times
\dots \times s_n \rightarrow s_0$ with $s_i \in \sorts$ for all $i \leq
n$. Function symbols of arity 0 are called \emph{constant
  symbols}. Similarly, predicate symbols $P \in \preds$ have an arity
$n \geq 0$ and sort $s_1 \times \dots \times s_n$. We assume dedicated
equality symbols $\equal_s \, \in \preds$ with the sort $s \times s$
for all sorts $s \in \sorts$, though we typically drop the explicit
subscript.
Terms are built from the function symbols in $\funs$ and (sorted)
variables taken from a countably infinite set $X$ that is disjoint
from $\funs$.  We denote by $t:s$ that term $t$ has sort $s$.
%
%

A $\sig$-atom $A$ is of the form $P(t_1,\dots,t_n)$ where $P \in
\preds$ is a predicate symbol of sort $s_1 \times \dots \times s_n$
and the $t_i$ are terms with $t_i: s_i$. A $\sig$-\emph{formula} $F$
is either a $\sig$-atom $A$, $\lnot F_1$, $F_1 \land
F_2$, $F_1 \lor F_2$, or $\forall x : s. F_1$ where $F_1$ and $F_2$ are
$\sig$-formulas. 
A $\sig$-\emph{literal} $L$ is either $A$ or $\lnot
A$ for a $\sig$-atom $A$.  A $\sig$-\emph{clause} $\clause$ is a
disjunction of $\sig$-literals.
A $\sig$-term, atom, or formula is said to be \emph{ground}, if no
variable appears in it.  For a set of formulas $\mathcal{K}$, we denote by
$\subterms(\mathcal{K})$ the set of all ground subterms that appear in
$\mathcal{K}$.

A $\sig$-sentence is a $\sig$-formula with no
free variables where the free variables of a formula are defined in
the standard fashion.
We typically omit $\sig$ if it is clear from the context.


\paragraph{Structures.}

Given a signature $\sig=(\sorts,\funs, \preds)$, a \emph{$\sig$-structure} $\alg$
is a function that maps each sort $s \in \sorts$ to a non-empty set $\alg(s)$,
each function symbol $f \in \funs$ of sort $s_1 \times \dots
\times s_n \rightarrow s_0$ to a function $\alg(f): \alg(s_1)
\times \dots \times \alg(s_n) \to \alg(s_0)$, and each predicate
symbol $P \in \preds$ of sort $s_1 \times \dots
\times s_n$ to a relation $\alg(s_1)
\times \dots \times \alg(s_n)$.  
We assume that all structures $\alg$ interpret each symbol $\equal_s$
by the equality relation on $\alg(s)$. For a $\sig$-structure $\alg$
where $\sig$ extends a signature $\sig_0$ with additional sorts and
function symbols, we write $\restrict{\alg}{\sig_0}$ for the
$\sig_0$-structure obtained by restricting $\alg$ to $\sig_0$.

Given a structure $\alg$ and a \emph{variable assignment}
$\varassign : X \rightarrow \alg$, the evaluation $t^{\alg,\varassign}$ of a
term $t$ in $\alg,\varassign$ is defined as usual.
%
For a structure $\alg$ and an atom $A$ of the form $P(t_1,\dots,t_n)$,
$(\alg, \varassign)$ satisfies $A$ iff $(t_1^{\alg,\varassign}, \dots,
t_n^{\alg,\varassign}) \in \alg(P)$.
This is written as $(\alg, \varassign) \models A$.
From this satisfaction relation of atoms and  $\sig$-structures,
we can derive the standard notions of the satisfiability of a formula,
satisfying a set of formulas $(\alg, \varassign) \models \{F_i\}$,
validity $\models F$, and entailment $F_1 \models F_2$. If a
$\sig$-structure $\alg$ satisfies a $\sig$-sentence $F$, we call
$\alg$ a model of $F$.

\paragraph{Theories and theory extensions.}
\label{sec:prelim:theory}
A \emph{theory} $\theory$ over signature $\sig$ is a set of $\sig$-structures.
We call a $\sig$-sentence $\axiom$ an \emph{axiom} if it is
the universal closure of a $\sig$-clause,
and we denote a set of $\sig$-axioms as $\axioms$.
We consider theories $\theory$ defined as a class of $\sig$-structures
that are models of a given set of $\sig$-sentences $\axioms$.
%
%
%
%

Let $\sig_0=(\sorts_0,\funs_0,\preds)$ be a signature and assume that
the signature
$\sig_1=(\sorts_0 \cup \sorts_e,\funs_0 \cup \funs_e, \preds)$ extends $\sig_0$ by new
sorts $\sorts_e$ and function symbols $\funs_e$.
We call the elements of $\funs_e$ \emph{extension symbols}
and terms starting with extension symbols \emph{extension terms}.
Given a $\sig_0$-theory $\theory_0$ and $\sig_1$-axioms $\extaxioms$,
we call $(\theory_0,\extaxioms, \theory_1)$ the \emph{theory
  extension} of $\theory_0$ with $\extaxioms$, where $\theory_1$ is
the set of all $\sig_1$-structures $\alg$ that are models of
$\extaxioms$ and whose reducts $\restrict{\alg}{\sig_0}$ are in
$\theory_0$. We often identify the theory extension with the theory
$\theory_1$.

\section{Problem}
We formally define the problem of satisfiability modulo theory and the
notion of local theory extensions in this section.

Let $\theory$ be a theory over signature $\sig$. Given a
$\sig$-formula $\aformula$, we say $\aformula$ is satisfiable modulo
$\theory$ if there exists a structure $M$ in $\theory$ and an
assignment $\varassign$ of the variables in $\aformula$ such that $(M,
\varassign) \models \aformula$. We define the ground
satisfiability modulo theory problem as the corresponding decision
problem for quantifier-free formulas.
\begin{problem}[Ground satisfiability problem for $\sig$-theory $\theory$]
\label{prob:smt}
\begin{description}
\item[input:] A quantifier-free $\sig$-formula $\aformula$. 
\item[output:] $\sat$ if $\aformula$ is satisfiable modulo $\theory$, $\unsat$ otherwise.
\end{description}
\end{problem}
We say the satisfiability problem for $\theory$ is \emph{decidable} if
there exists a procedure for the above problem which always terminates
with $\sat$ or $\unsat$.
We write entailment modulo a theory as $\phi \models_\theory \psi$.

We say an axiom of a theory extension is \emph{linear} if all the
variables occur under at most one extension term.
We say it is \emph{flat} if there there is no nesting of terms
containing variables.
It is easy to linearize and flatten the axioms by using additional
variables and equality.
As an example, $\forall x. \phi$ with $f(x)$
and $f(g(x))$ as terms in $F$ may be written as
\[\forall x y z. x \equal y \wedge z \equal g(y) \implies F' \]
where $F'$ is obtained from $F$ by replacing $f(g(x))$ with $f(z)$.
For the remainder of the paper, we assume that all extension axioms
$\extaxioms$ are flat and linear. For the simplicity of the presentation,
we assume that if a variable appears below a function symbol
then that symbol must be an extension symbol.

\begin{definition}[Local theory extensions]
\label{def:lte}
A theory extension $(\basetheory, \extaxioms, \exttheory)$ is
\emph{local} if for any set of ground $\Sigma_1$-literals $G$: $G$ is
satisfiable modulo $\exttheory$ if and only if $G \union \extaxioms[G]$
is satisfiable modulo $\basetheory$ extended with free function
symbols. Here $\extaxioms[G]$ is the set of instances of $\extaxioms$
where the subterms of the instantiation are all subterms of $G$ or
$\extaxioms$ (in other words, they do not introduce new terms).
\end{definition}
For simplicity, in the rest of this paper, we work with theories
$\basetheory$ which have decision procedures for not just
$\basetheory$ but also $\basetheory$ extended with free function
symbols. Thus, we sometimes talk of satisfiability of a
$\Sigma_1$-formula with respect a $\Sigma_0$-theory $\basetheory$, to
mean satisfiability in the $\basetheory$ with the extension symbols in
$\Sigma_1$ treated as free function symbols. In terms of SMT, we only
talk of extensions of theories containing uninterpreted functions
(\texttt{UF}).

A naive decision procedure for ground SMT of a local theory extension
$\exttheory$ is thus to generate all possible instances of the axioms
$\extaxioms$ that do not introduce new ground terms, thereby reducing
to the ground SMT problem of $\basetheory$ extended with free functions.

\medskip \noindent
\emph{Hierarchical extensions.}
Note that local theory extensions can be stacked to form hierarchies
\[((\dots((\theory_0, \mathcal{K}_1, \theory_1), \mathcal{K}_2,
\theory_2),\dots), \mathcal{K}_n, \theory_n).\]
Such a hierarchical
arrangement of extension axioms is often useful to modularize locality
proofs. In such cases, the condition that variables are only allowed
to occur below extension symbols (of the current extension) can be
relaxed to any extension symbol of the current level or below. The
resulting theory extension can be decided by composing 
procedures for the individual extensions. Alternatively, one can use a
monolithic decision procedure for the resulting theory $\theory_n$, which can
also be viewed as a single local theory extension $(\theory_0,
\mathcal{K}_1 \cup \dots \cup \mathcal{K}_n, \theory_n)$. In our
experimental evaluation, which involved hierarchical extensions, we followed the latter approach.



\section{Algorithm}
In this section, we describe a decision procedure for a local theory
extension, say $(\basetheory, \extaxioms, \exttheory)$, which
can be easily implemented in most SMT solvers with quantifier
instantiation support.  We describe our procedure
$\theoryproc_{\exttheory}$ as a theory module in a typical SMT solver
architecture. For simplicity, we separate out the interaction between
theory solver and core SMT solver. We describe the procedure abstractly as
taking as input:
\begin{itemize}
\item
 the original formula $\aformula$,
\item 
a set of extension axioms $\extaxioms$,
\item
a set of instantiations of axioms that have already been
made, $Z$, and
\item
a set of $\basetheory$ satisfiable ground literals $G$ such that 
$G \models \aformula \wedge (\bigwedge_{\psi \in Z} \psi)$, and
\item a set equalities $E \subseteq G$ between terms.
\end{itemize}
It either returns
\begin{itemize}
\item
 $\sat$, denoting that $G$ is $\exttheory$ satisfiable; or
\item 
a new set of instantiations of the axioms, $Z'$.
\end{itemize}

For completeness, we describe briefly the way we envisage the
interaction mechanism of this module in a DPLL(T) SMT solver. Let the
input problem be $\aformula$. The SAT solver along with the theory
solvers for $\basetheory$ will find a subset of literals $G$ from
$\aformula \wedge (\bigwedge_{\psi \in Z} \psi)$ such that its
conjunction is satisfiable modulo $\basetheory$. If no such satisfying
assignment exists, the SMT solver stops with $\unsat$. One can think
of $G$ as being simply the literals in $\aformula$ on the SAT solver
trail. $G$ will be sent to $\theoryproc_{\exttheory}$ along with
information known about equalities between terms. The set $Z$ can be
also thought of as internal state maintained by the
$\exttheory$-theory solver module, with new instances $Z'$ sent out as
theory lemmas and $Z$ updated to $Z \cup Z'$ after each call to
$\theoryproc_{\exttheory}$. If $\theoryproc_{\exttheory}$ returns
$\sat$, so does the SMT solver and stops. On the other hand, if it
returns a new set of instances, the SMT solver continues the search to
additionally satisfy these.

\medskip \noindent
\emph{E-matching.}
In order to describe our procedure, we introduce the well-studied
E-match\-ing problem. Given a universally quantified $\Sigma$-sentence
$K$, let $X(K)$ denote the quantified variables. Define a
$\sig$-substitution $\sub$ of $K$ to be a mapping from variables
$X(K)$ to $\sig$-terms of corresponding sort. Given a $\sig$-term $p$,
let $p\sub$ denote the term obtained by substituting variables in $p$
by the substitutions provided in $\sub$. Two substitutions $\sub_1$,
$\sub_2$ with the same domain $X$ are equivalent modulo a set of
equalities $E$ if $\forall x \in X.\, E \models \sub_1(x) \equal
\sub_2(x)$. We denote this as $\sub_1 \sim_E \sub_2$.
%
\begin{problem}[E-matching]
\begin{description}
\item[input:] A set of ground equalities $\equalities$, a set of
  $\sig$-terms $\groundterms$, and patterns $\Patterns$.
\item[output:] The set of substitutions $\sub$ over the variables in
  $p$, modulo $E$, such that for all $p \in \Patterns$ there exists a
  $t \in G$ with $E \models t \equal p\sub$.
\end{description}
\end{problem}
E-matching is a well-studied problem, specifically in the context of
SMT. An algorithm for E-matching that is efficient and backtrackable
is described in \cite{MB07}. We denote this procedure by
$\ematchproc$.


\begin{figure}[t]
$\theoryproc_{\exttheory}(\aformula, \extaxioms, Z, G, E)$

Local variable: $Z'$, initially an empty set.
\begin{enumerate}
\item For each $K \in \extaxioms$:
\begin{enumerate}
\item Define the set of patterns $P$ to be the function symbols in $K$
  containing variables. We observe that because the axioms are linear
  and flat, these patterns are always of the form $f(x_1,\dots,x_n)$
  where $f$ is an extension symbol and the $x_i$ are quantified
  variables.
\item Run $\ematchproc(E, G, P)$ obtaining substitutions
  $\subs$. Without loss of generality, assume that $\sub \in \subs$
  returned by the algorithm are such that $\subterms(K\sub) \subseteq
  \subterms(G \union \extaxioms)$. For the special case of the
  patterns in (a), for any $\sub$ not respecting the condition there
  exists one in the equivalence class that respects the
  condition. \knew{Formally, $\forall \sub. \exists \sub'.  \sub' \sim_E
  \sub \wedge \subterms(K\sub') \subseteq \subterms(G \union
  \extaxioms)$.} We make this assumption only for simplicity of
  arguments later in the paper. \knew{If one uses an E-matching procedure
  not respecting this constraint, our procedure will still be
  terminating and correct (albeit total number of instantiations suboptimal).}
  \kold{Since the e-matching algorithm is required to return substitution
  modulo $\sim_E$, for simplicity of arguments later in the paper we
  make this assumption.}

\item For each $\sub \in \subs$, if there exists no $K\sub'$ in $Z$
  such that $\sub \sim_E \sub'$, then add $K\sub$ to $Z'$ as a new
  instantiation to be made.
\end{enumerate}
\item If $Z'$ is empty, return $\sat$, else return $Z'$.
\end{enumerate}
\caption{Procedure $\theoryproc_{\exttheory}$}
\label{fig:algo}
\end{figure}

The procedure $\theoryproc_{\exttheory}(\aformula, \extaxioms, Z, G,
E)$ is given in Fig.~\ref{fig:algo}. Intuitively, it 
adds all the new instances along the current search path that
are required for local theory reasoning as given in Definition
\ref{def:lte}, but modulo equality. For each axiom $K$ in
$\extaxioms$, the algorithm looks for function symbols containing
variables. For example, if we think of the monotonicity axiom in
Sect. \ref{sec:example}, these would be the terms $f(x)$ and
$f(y)$. These terms serve as patterns for the E-matching
procedure. Next, with the help of the E-matching algorithm, all
\emph{new} instances are computed (to be more precise, all instances
for the axiom $K$ in $Z$ which are equivalent modulo $\sim_E$ are
skipped). If there are no new instances for any axiom in $\extaxioms$,
and the set $G$ of literals implies $\aformula$, we stop with \textsf{sat}. as
effectively we have that $G \union \extaxioms[G]$ is satisfiable
modulo $\basetheory$. Otherwise, we return this set.

We note that though the algorithm $\theoryproc_{\exttheory}$ may
\emph{look} inefficient because of the presence of nested loops,
keeping track of which substitutions have already happened, and which
substitutions are new. However, in actual implementations all of this
is taken care of by the E-matching algorithm. There
has been significant research on fast, incremental algorithms for
E-matching in the context of SMT, and one advantage of our approach is
to be able to leverage this work.




\medskip \noindent
\emph{Correctness.}
The correctness argument relies on two aspects: one, that if the SMT
solver says $\sat$ (resp. $\unsat$) then $\aformula$ is
satisfiable (resp. unsatisfiable) modulo $\exttheory$, and second,
that it terminates.

%
For the case where the output is $\unsat$, the correctness follows
from the fact that $Z$ only contains instances of $\extaxioms$. The
$\sat$ case is more tricky, but the main idea is that the set of
instances made by $\theoryproc_{\exttheory}(\aformula, \extaxioms, Z,
G, E)$ are logically equivalent to $\extaxioms[G]$. Thus, when the
solver stops, $G \union \extaxioms[G]$ is satisfiable modulo
$\basetheory$. As a consequence, $G$ is satisfiable modulo
$\exttheory$. Since $G \models \aformula$, we have that
$\aformula$ is satisfiable modulo $\exttheory$.

The termination relies on the fact that the instantiations returned by
procedure
$\theoryproc_{\exttheory}(\aformula, \extaxioms, Z, G, E)$ do not add
new terms, and they are always a subset of
$\extaxioms[\aformula]$. Since, $\extaxioms[\aformula]$ is finite,
eventually $\theoryproc$ will stop making new instantiations. Assuming
that we have a terminating decision procedure for the ground SMT problem
of $\basetheory$, we get a terminating decision procedure for
$\exttheory$.



\begin{theorem}
  An SMT solver with theory module $\theoryproc_{\exttheory}$ is a
  decision procedure for the satisfiability problem modulo $\exttheory$.
\end{theorem}

\paragraph{Psi-local theories.}
We briefly explain how our approach can be extended to the more
general notion of Psi-local theory
extensions~\cite{IhlemannETAL08LocalReasoninginVerification}.
Sometimes, it is not sufficient to consider only local instances of
extension axioms to decide satisfiability modulo a theory
extension. For example, consider the following set of ground literals:
\[G = \{f(a) = f(b), a \neq b\}
\] 
Suppose we interpret $G$ in a theory of an injective function $f: S
\to S$ with a partial inverse $g: S \to S$ for some set $S$. We can
axiomatize this theory as a theory extension of the theory of
uninterpreted functions using the axiom
\[K = \forall x, y.\, f(x) = y \implies g(y) = x \enspace.
\] 
$G$ is unsatisfiable in the theory extension, but the local instances
of $K$ with respect to the ground terms $\subterms(G)=\{a,b,f(a),f(b)\}$ are
insufficient to yield a contradiction in the base theory. However, if
we consider the local instances with respect to the larger set of
ground terms 
\[\Psi(\subterms(G)) = \{a,b,f(a),f(b),g(f(a)),g(f(b))\},
\] 
then we obtain, among others, the instances
\[f(a) = f(b) \implies g(f(b)) = a \quad \text{ and } \quad f(b) = f(a)
\implies g(f(a)) = b \enspace. \]
Together with $G$, these instances are unsatisfiable in the base theory.

The set $\Psi(\subterms(G))$ is computed as follows:
\[ \Psi(\subterms(G)) = \subterms(G) \cup \pset{g(f(t))}{t \in \subterms(G)}\] It turns
out that considering local instances with respect to $\Psi(\subterms(G))$
is sufficient to check satisfiability modulo the theory extension $K$
for arbitrary sets of ground clauses $G$. Moreover, $\Psi(\subterms(G))$ is
always finite. Thus, we still obtain a decision procedure
for the theory extension via finite instantiation of extension axioms.
Psi-local theory extensions formalize this idea. In particular, if
$\Psi$ satisfies certain properties including monotonicity and
idempotence, one can again provide a model-theoretic characterization
of completeness in terms of embeddings of partial models. We refer the
reader to~\cite{IhlemannETAL08LocalReasoninginVerification} for the
technical details.

To use our algorithm for deciding satisfiability of a set of ground
literals $G$ modulo a Psi-local theory extension
$(\basetheory,\extaxioms,\exttheory)$, we simply need to add an
additional preprocessing step in which we compute $\Psi(\subterms(G))$ and
define
$G' = G \cup \pset{\mathtt{instclosure}(t)}{t \in \Psi(\subterms(G))}$
where $\mathtt{instclosure}$ is a fresh predicate symbol. Then calling
our procedure for $\exttheory$ with $G'$ decides satisfiability of $G$
modulo $\exttheory$.

\kold{
One heuristic for Psi-local theories that we found to improve the
performance of our algorithm in our experiments is as follows:
E-matching with the terms that are in $\Psi(\subterms(G))$ but not in
$\subterms(G)$ should be delayed. That is, only after an auxiliary term $t
\in \Psi(\subterms(G)) \setminus \subterms(G)$ occurs in an instance of an axiom
that has been generated in a previous instantiation round, is $t$
allowed to be used in substitutions that generate additional
instances.}




\section{Implementation and Experimental Results}

\smartparagraph{Benchmarks.}
We evaluated our techniques on a set of benchmarks
generated by the deductive verification tool
\tool{GRASShopper}~\cite{grasshopper-tool}. The benchmarks encode
memory safety and functional correctness properties of programs that
manipulate complex heap-allo\-cated data structures. The programs are
written in a type-safe imperative language without garbage
collection. The tool makes no simplifying assumptions about these
programs like acyclicity of heap structures.

\tool{GRASShopper} supports mixed specifications in (classical) first-order
logic and separation logic
(SL)~\cite{Reynolds02SeparationLogic}. The tool reduces the program and specification to verification
conditions that are encoded in a hierarchical
combination of (Psi-)local theory extensions. This hierarchy of
extensions is organized as follows:
\begin{enumerate}
\item \textit{Base theory:} at the lowest level we have
  \texttt{UFLIA}, the theory of uninterpreted functions and linear
  integer arithmetic, which is directly supported by SMT solvers.
\item \textit{GRASS:} the first extension layer consists of the theory
  of graph reachability and stratified sets. This theory is a disjoint
  combination of two local theory extensions: the theory of linked
  lists with reachability~\cite{DBLP:conf/popl/LahiriQ08} and the theory of sets
  over interpreted elements~\cite{DBLP:conf/birthday/Zarba03}.
\item \textit{Frame axioms:} the second extension layer consists of
  axioms that encode the frame rule of separation logic. This theory
  extension includes arrays as a subtheory.
\item \textit{Program-specific extensions:} The final extension layer
  consists of a combination of local extensions that encode properties
  specific to the program and data structures under
  consideration. These include:
  \begin{itemize}
  \item axioms defining memory footprints of SL specifications,
  \item axioms defining structural constraints on the shape of data structures,
  \item sorted constraints, and
  \item axioms defining partial inverses of certain functions, e.g.,
    to express injectivity of functions and to specify the content of
    data structures.
  \end{itemize}
\end{enumerate}
We refer the interested reader to~\cite{DBLP:conf/cav/PiskacWZ13,
  grasshopper, DBLP:conf/cav/PiskacWZ14} for further details about the
encoding.

The programs considered include sorting algorithms, common data
structure operations, such as inserting and removing elements, as well
as complex operations on abstract data types. Our selection of data
structures consists of singly and doubly-linked lists, sorted lists,
nested linked lists with head pointers, binary search trees, skew
heaps, and a union find data structure. The input programs comprise
108 procedures with a total of 2000 lines of code, 260 lines of
procedure contracts and loop invariants, and 250 lines of data
structure specifications (including some duplicate specifications that
could be shared across data structures).
\knew{The verification of these specifications are reduced by
  \tool{GRASShopper} to 816 SMT queries, each serves as one benchmark
  in our experiments.}\kold{The total benchmark set comprises 816
  benchmarks, each consisting of a single SMT query.} 802 benchmarks
are unsatisfiable. The remaining 14 satisfiable benchmarks stem from
programs that have bugs in their implementation or specification. All
of these are genuine bugs that users of \tool{GRASShopper} made while
writing the
programs.\footnote{See~\url{www.cs.nyu.edu/~kshitij/localtheories/}
  for the programs and benchmarks used.}
We considered several versions of each benchmark, which we describe in
more detail below. Each of these versions is encoded as an SMT-LIB 2
input file.

\smartparagraph{Experimental setup.}
All experiments were conducted on the StarExec
platform~\cite{StumpST14} with a CPU time limit of one hour and a
memory limit of 100 GB.  We focus on the SMT solvers
\tool{CVC4}\cite{conf/cav/BarrettCDHJKRT11} and
\tool{Z3}\cite{MouraBjoerner08Z3}\footnote{We used the version of \tool{Z3} 
  downloaded from the git master branch at \url{http://z3.codeplex.com} on Jan 17,
2015.} as both support \texttt{UFLIA} and
quantifiers via E-matching.  This version of \tool{CVC4} is a fork of
v1.4 with special support for quantifiers.\footnote{This version is
  available at \url{www.github.com/kbansal/CVC4/tree/cav14-lte-draft}.  }

\kold{
We instrumented GRASShopper to eagerly instantiate
all of the (Psi-)local theory axioms (modulo top level equalities).
The resulting ground SMT query is satisfiable if and only if the original query
is satisfiable in the local theory extension.
We also instrumented GRASShopper to generate benchmarks where
(Psi-)local theory axioms are provided as quantified formulas.  We
call these \emph{uninstantiated} benchmarks.
}

In order to be able to test our approach with both \tool{CVC4} and
\tool{Z3}, wherever possible we transformed the benchmarks to simulate
our algorithm. We describe these transformations in this paragraph.
First, the quantified formulas in the benchmarks were linearized and
flattened, and annotated with patterns to simulate Step 1(a) of our
algorithm (this was done by \tool{GRASShopper} in our experiments, but
may also be handled by an SMT solver aware of local theories). Both
\tool{CVC4} and \tool{Z3} support using these annotations for controlling
instantiations in their E-matching procedures.
In order to handle Psi-local theories, the additional terms required
for completeness were provided as dummy assertions, so that these
appear as ground terms to the solver. In \tool{CVC4}, we also made some
changes internally so as to treat these assertions specially and apply
certain additional optimizations which we describe later in this section.

\smartparagraph{Experiment 1.}
\begin{figure}[t]
  \begin{center}
    \subfloat[\tool{CVC4} with baseline algorithm]{\label{fig:instantiations:a}
      \includegraphics[width=0.48\textwidth]{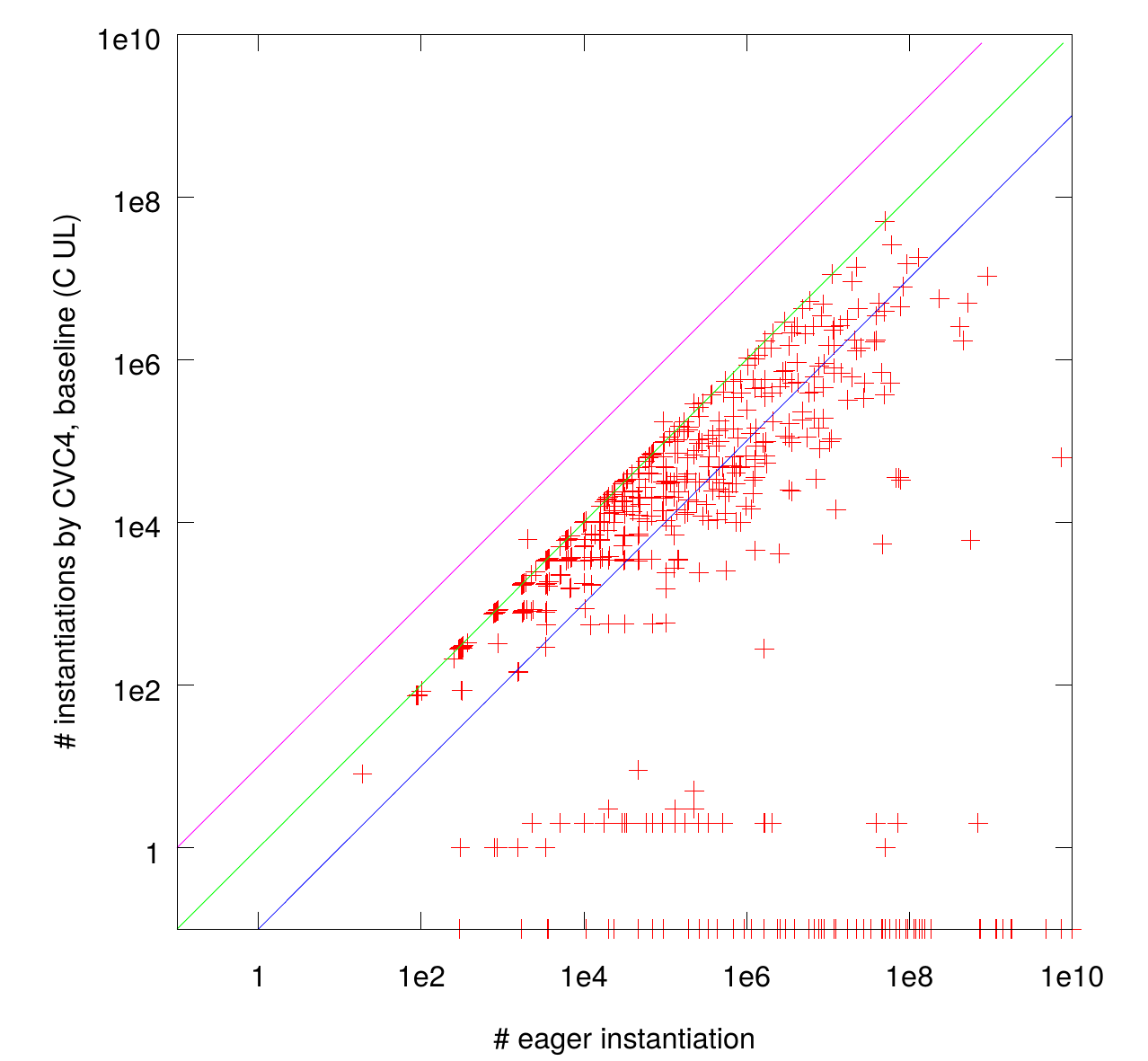}
    }
    \subfloat[\tool{CVC4} with optimized algorithm]{\label{fig:instantiations:b}
      \includegraphics[width=0.48\textwidth]{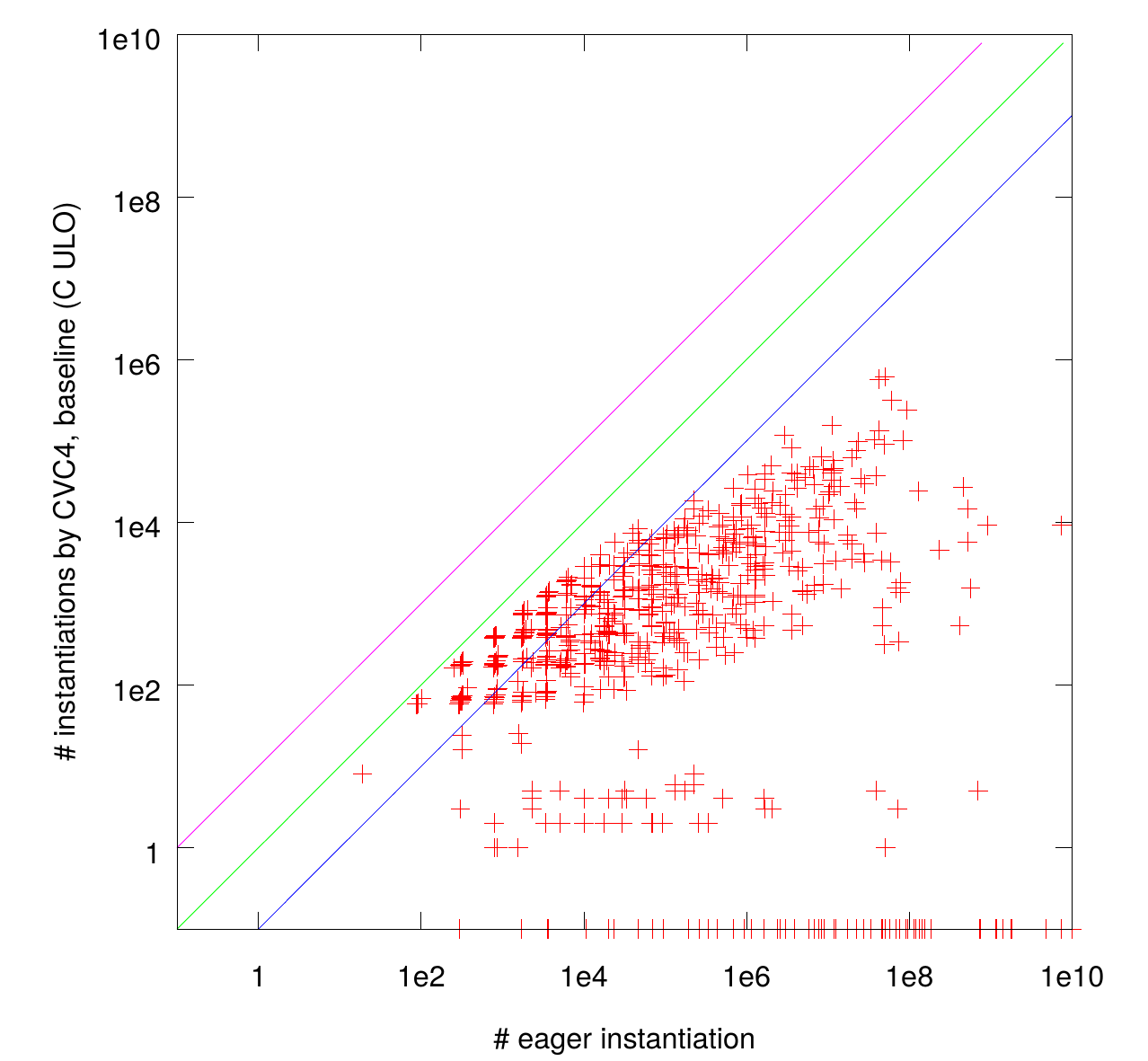}
    }
  \end{center}
  \vspace*{-1em}
  \caption{\# of eager instantiations vs. E-matching instantiations
    inside the solver}\label{fig:instantiations}
  \vspace*{1em}
\end{figure}

Our first experiment aims at comparing the effectiveness of eager
instantiation versus incremental instantiation up to congruence (as
done by E-matching).
Figure~\ref{fig:instantiations} charts the number of eager
instantiations versus the number of E-matching instantiations for each
query in a logarithmic plot.\footnote{Figure~\ref{fig:instantiations}
  does not include timeouts for \tool{CVC4}.}
%
Points lying on  the central line have an equal number of instantiations in both
series while points lying on the lower line have 10 times as many eager
instantiations as E-matching instantiations.
(The upper line corresponds to $\frac{1}{10}$.)
Most benchmarks require substantially more eager instantiations.
We instrumented \tool{GRASShopper} to eagerly instantiate
all axioms.
Subfigure (a) compares upfront instantiations with a baseline
implementation of our E-matching algorithm. Points along the $x$-axis
required no instantiations in \tool{CVC4} to conclude unsat.
We have plotted the above charts up to 10e10 instantiations.
There were four outlying benchmarks where upfront instantiations had between
10e10 and up to 10e14 instances. E-matching had zero instantiations for all four.
Subfigure (b) compares against an optimized version of our algorithm
implemented in \tool{CVC4}. It shows that incremental solving reduces the
number of instantiations significantly, often by several orders of
magnitude. The details of these optimizations are given later in the
section.

\smartparagraph{Experiment 2.}
\begin{table}[t]
  \begin{center}
    {\small
    \begin{tabular}{|\ColumnSpace lr \ColumnBar rr \ColumnBar rr \ColumnBar rr \ColumnBar rr \ColumnBar rr \ColumnBar rr  \ColumnSpace |}
      \hline
      & &
      \abbcvc & \abbnoinst\abbdefault &
      \abbcvc & \abbnoinst\abblocal &
      \abbcvc & \abbnoinst\abblocal\abbopt &
      \abbz   & \abbnoinst\abbdefault &     
      \abbz   & \abbnoinst\abblocal &       
      \abbz   & \abbnoinst\abblocal\abbopt
      \\
      family  & \# & \# &  time & \# &  time & \# & time & \# & time & \# & time & \# & time \\ \hline
      sl lists & 139 & 127 & 70 & 139 & 383 & \textbf{139} & \textbf{17} & 138 & 1955 & 138 & 1950 & 139 & 68 \\
      dl lists & 70 & 66 & 1717 & 70 & 843 & \textbf{70} & \textbf{33} & 56 & 11375 & 56 & 11358 & 70 & 2555 \\
      sl nested & 63 & 63 & 1060 & 63 & 307 & \textbf{63} & \textbf{13} & 52 & 6999 & 52 & 6982 & 59 & 1992 \\
      sls lists & 208 & 181 & 6046 & 204 & 11230 & \textbf{208} & \textbf{3401} & 182 & 20596 & 182 & 20354 & 207 & 4486 \\
      trees & 243 & 229 & 2121 & 228 & 22042 & \textbf{239} & \textbf{7187} & 183 & 41208 & 183 & 40619 & 236 & 27095 \\
      soundness & 79 & 76 & 17 & 79 & 1533 & \textbf{79} & \textbf{70} & 76 & 7996 & 76 & 8000 & 79 & 336 \\
      \hline
      sat & 14 & - & - & 14 & 670 & \textbf{14} & \textbf{12} & - & -& 10 & 3964 & 14 & 898 \\
      \hline
      total & 816 & 742 & 11032 & 797 & 37009 &\textbf{ 812} & \textbf{10732} & 687 & 90130 & 697 & 93228 & 804 & 37430 \\
      \hline
    \end{tabular}
    }
  \end{center}
  \caption{Comparison of solvers on
    uninstantiated benchmarks (time in
    sec.)}\label{table:cvc4:comp:noinst}
\end{table}

Next, we did a more thorough comparison on running times and number of
benchmarks solved for \emph{uninstantiated benchmarks}. These results
are in Table~\ref{table:cvc4:comp:noinst}.
The benchmarks are partitioned according to
the types of data structures occurring in the programs from which the
benchmarks have been generated. Here, ``sl'' stands for singly-linked, ``dl''
for double-linked, and ``sls'' for sorted singly-linked. The binary search
tree, skew heap, and union find benchmarks have all been summarized in
the ``trees'' row. The row ``soundness'' contains unsatisfiable
benchmarks that come from programs with incorrect code or
specifications. These programs manipulate various types of data
structures. The actual satisfiable queries that reveal the bugs in
these programs are summarized in the ``sat'' row.

We simulated our algorithm and ran these experiments on both
\tool{CVC4} (C) and \tool{Z3} obtaining similar improvements with
both. We ran each with three configurations:
\begin{description}
\item[\abbnoinst\abbdefault] Default. For comparison purposes, we ran the solvers with
  default options. \tool{CVC4}'s default solver uses an E-matching
  based heuristic instantiation procedure, whereas \tool{Z3}'s uses
  both E-matching and model-based quantifier instantiation (MBQI). For
  both of the solvers, the default procedures are incomplete for our
  benchmarks.
\item[\abbnoinst\abblocal] These columns refer to the E-matching based complete
  procedure for local theory extensions (algorithm in
  Fig.~\ref{fig:algo}).\footnote{
    The configuration \abbcvc\xspace \abbnoinst\abblocal\xspace
    had one memory out on a benchmark in the tree family.
}
\item[\abbnoinst\abblocal\abbopt] Doing instantiations inside the solver instead of upfront,
  opens the room for optimizations wherein one tries some
  instantiations before others, or reduces the number of
  instantiations using other heuristics that do not affect
  completeness. The results in these columns show the effect of all such
  optimizations.
\end{description} 
As noted above, the \abbnoinst\abblocal\xspace and \abbnoinst\abblocal\abbopt\xspace
procedures are both complete, whereas \abbnoinst\abbdefault\xspace is
not. This is also reflected in the ``sat'' row in
Table~\ref{table:cvc4:comp:noinst}. Incomplete Instantiation-based
procedures cannot hope to answer ``sat''. A significant improvement can be seen
between the \abbnoinst\abblocal\xspace and \abbnoinst\abblocal\abbopt\xspace
columns.
\knew{The general thrust of the optimizations was to avoid blowup of
  instantiations by doing ground theory checks on a subset of
  instantiations. Our intuition is that the theory lemmas learned from
  these checks eliminate large parts of the search space before we do
  further instantiations.}
\kold{Some of the optimizations we found helpful on these
benchmarks are as follows (most of these seem helpful in general):}

\knew{For example, we used a heuristic for Psi-local theories inspired
  from the observation that the axioms involving Psi-terms are needed
  mostly for completeness, and that we can prove
  unsatisfiable without instantiating axioms with these terms most of the time.}
%
\kold{As noted above, our benchmarks had some Psi-local theory
  extensions.} We tried an approach where the instantiations were
  staged. First, the instantiations were done according to the
  algorithm in Fig.~\ref{fig:algo} for locality with respect to ground terms from
  the original query. Only when those were saturated, the
  instantiations for the auxiliary Psi-terms were generated. We 
  found this to be very helpful. Since this required non-trivial changes
  inside the solver, we only implemented this optimization in
  \tool{CVC4}; but we think that staging instantiations for Psi-local
  theories is a good strategy in general.

  \knew{A second optimization, again with the idea of cutting
  instantiations,} was adding assertions in the benchmarks of the form
  $(a = b) \vee (a \neq b)$ where $a$, $b$ are ground terms. This forces an arbitrary
  arrangement over the ground terms before the
  instantiation procedure kicks in. \knew{
    Intuitively, the solver first does
checks with many terms equal to each other (and hence fewer
instantiations) eliminating as much of the search space as
possible. Only when equality or disequality is relevant to the
reasoning is the solver forced to instantiate with terms disequal
to each other.}
\kold{If many of the terms are equal, it
leads to fewer instantiations in the E-matching call on the current
trail.} \knew{One may contrast this with ideas being used
  successfully in the care-graph-based theory combination framework in
  SMT where one needs to try all possible arrangements of equalities
  over terms. It has been observed that equality or disequality is
  sometimes relevant only for a subset of pairs of terms. Whereas in
  theory combination this idea is used to cut down the number of
  arrangements that need to be considered, we use it to reduce the
  number of unnecessary instantiations.}
 We found this really helped \tool{CVC4} on many
 benchmarks.
\ksays{Sentence from rebuttal not incorporated:
 The optimizations which force equality between terms reduce polynomial
blowup in combination with E-matching. The intuition here is that 
}
 
 Another optimization was instantiating special cases of the
 axioms first by enforcing equalities between variables of the same
 sort, before doing a full instantiation. We did this for axioms that
 yield a particularly large number of instances (instantiations
 growing with the fourth power of the number of ground terms). Again,
 we believe this could be a good heuristic in general.

\smartparagraph{Experiment 3.}
\begin{table}[t]
  \begin{center}
    {\small
    \begin{tabular}{|\ColumnSpace lr \ColumnBar  rr \ColumnBar rr \ColumnBar rr \ColumnBar rr \ColumnBar rr  \ColumnSpace |}
      \hline
      & &
      \abbcvc & \abbinst\abblocal &
      \abbcvc & \abbinst\abblocal\abbopt &
      \abbz   & \abbinst\abbmbqi &     
      \abbz   & \abbinst\abblocal &       
      \abbz   & \abbinst\abblocal\abbopt
      \\
      family  & \# & \# &  time & \# &  time  & \# & time & \# & time & \# & time \\ \hline
      sl lists & 139 & 139 & 664 & 139 & 20 & \textbf{139} & \textbf{9} & 139 & 683 & 139 & 29 \\
      dl lists & 70 & 70 & 3352 & 70 & 50 & \textbf{70} & \textbf{41} & 67 & 12552 & 70 & 423 \\
      sl nested & 63 & 63 & 2819 & 63 & 427 & \textbf{63} & \textbf{182} & 56 & 7068 & 62 & 804 \\
      sls lists & 208 & 206 & 14222 & 207 & 3086 & \textbf{208} & \textbf{37} & 203 & 17245 & 208 & 1954 \\
      trees & 243 & 232 & 7185 & 243 & 6558 & \textbf{243} & \textbf{663} & 222 & 34519 & 242 & 8089 \\
      soundness & 79 & 78 & 156 & 79 & 49 & \textbf{79} & \textbf{23} & 79 & 2781 & 79 & 39 \\
      \hline
      sat & 14 & 14 & 85 & \textbf{14} & \textbf{22} & 13 & 21 & 12 & 1329 & 14 & 109 \\
      \hline
      total & 816 & 802 & 28484 &\textbf{ 815} & 10213 &\textbf{ 815} & \textbf{976} & 778 & 76177 & 814 & 11447 \\
      \hline
    \end{tabular}
    }
  \end{center}
  \caption{Comparison of solvers on partially instantiated benchmarks (time in sec.)}\label{table:cvc4:comp:inst}
\end{table}

\knew{ Effective propositional Logic (EPR)
 is the fragment of first order-logic consisting of formulas of the shape
 $\exists\vec{x}\forall\vec{y}.G$ with $G$ quantifier-free and where
 none of the universally quantified variables $\vec{y}$ appears below
 a function symbol in $G$. Theory extensions that fall into EPR are
 always local. Our third exploration is to see if we can exploit
 dedicated procedures for this fragment when such fragments occur in
 the benchmarks.}
\ksays{I think we don't really define EPR anywhere properly or why it is a LTE. That explains the confusion we had in the reviews. @Thomas: Any suggestions to briefly define/summarize, or it current text good enough?}
\kold{In the process of experimenting, we also noticed that most of the
instantiations were coming from axioms which fall into the EPR
fragment, or can be partially instantiated to fall into EPR.}
For the EPR fragment, \tool{Z3} has a complete decision procedure that
uses model-based quantifier instantiation. We therefore implemented a
hybrid approach wherein we did upfront partial instantiation to the EPR
fragment using E-matching with respect to top-level equalities \knew{(as
described in our algorithm). The resulting EPR benchmark is then
decided using Z3's MBQI mode.}
This approach can only be expected to help where there are
EPR-like axioms in the benchmarks, and we did have some which were
heavier on these.
We found that on \kold{the considered benchmarks}\knew{singly linked
  list and tree benchmarks} this hybrid algorithm significantly
outperforms all other solver configurations that we have tried in our
experiments\kold{ (except on the satisfiable benchmarks)}. \knew{On the
other hand, on nested list benchmarks, which make more heavy use of purely equational
axioms, this technique does not help compared to only
using E-matching because the partial instantiation already yields
ground formulas.}\kold{this suggests that more sophisticated combinations of
E-matching and MBQI that further increase performance can be realized
inside the solver while retaining completeness.}
\ksays{See Apx.~\ref{rebuttalnotes} for rebuttal text for this part}


The results with our hybrid algorithm are summarized in Column Z3 \abbinst\abbmbqi\xspace
of Table~\ref{table:cvc4:comp:inst}. Since EPR is a special case of
local theories, we also tried our E-matching based algorithm on these
benchmarks. We found that the staged instantiation improves
performance on these as well.  The optimization that help on the
uninstantiated benchmarks also work here. These results are 
summarized in the same table.

Overall, our experiments indicate that there is a lot of potential in
the design of quantifier modules to further improve the performance of
SMT solvers, and at the same time make them complete on more
expressive decidable fragments.

\section{Conclusion}

We have presented a new algorithm for deciding local theory
extensions, a class of theories that plays an important role in
verification applications. Our algorithm relies on existing SMT solver
technology so that it can be easily implemented in today's solvers. In
its simplest form, the algorithm does not require any modifications to
the solver itself but only trivial syntactic modifications to its
input. These are: (1) flattening and linearizing the extension axioms;
and (2) adding trigger annotations to encode locality constraints for
E-matching. In our evaluation we have experimented with different
configurations of two SMT solvers, implementing a number of
optimizations of our base line algorithm. Our results suggest
interesting directions to further improve the quantifier modules of
current SMT solvers, promising better performance and usability for
applications in automated verification.

%


\paragraph{Acknowledgements}
This work was supported
in part by the National Science Foundation under grants CNS-1228768 and
CCF-1320583, and by the \href{http://erc.europa.eu/}{European Research Council} under the
European Union's Seventh Framework Programme (FP/2007-2013) / ERC Grant
Agreement nr.~306595 \href{http://stator.imag.fr/}{\mbox{``STATOR''}}.

\bibliographystyle{plain}
\bibliography{biblio}

\ifDRAFT
\appendix
\input{proofappendix}
\fi
\end{document}